\documentclass[runningheads]{llncs}
\usepackage[T1]{fontenc}
\usepackage{pdfpages} 
\usepackage{booktabs}
\usepackage{xurl}
\usepackage{hyperref}
\usepackage[table]{xcolor}
\usepackage{array}

\usepackage{placeins}    
\usepackage{float}       

\usepackage{longtable}
\usepackage{multirow}

 \usepackage{multirow}
 \usepackage[table,xcdraw]{xcolor}
 \usepackage[normalem]{ulem}
 \useunder{\uline}{\ul}{}
 \usepackage{longtable}
\usepackage{graphicx}
\begin{document}
\title{AI-Generated Rubric Interfaces: K–12 Teachers’ Perceptions and Practices}
%
%

\author{
Bahare Riahi\inst{1}\orcidID{0009-0005-4560-4857}\and
Sayali Patukale\inst{1}\orcidID{0000-0002-2960-9915}\and
Joy Niranjan\inst{1}\orcidID{0009-0006-1724-7271}
\and
Yogya Koneru\inst{1}\and
Tiffany Barnes\inst{1}\orcidID{0000-0002-6500-9976}\and
Veronica Cateté\inst{1}\orcidID{0000-0002-7620-7708}
}

\authorrunning{B. Riahi et al.}
%
\institute{North Carolina State University, Raleigh NC 27595, USA \\
\email{\{briahi, spatuka, sniranj2, ykoneru, vmcatete, tmbarnes\}@ncsu.edu}}

\maketitle              
\begin{abstract}
This study investigates K–12 teachers’ perceptions and experiences with AI-supported rubric generation during a summer professional development workshop (n = 25). Teachers used MagicSchool.ai to generate rubrics and practiced prompting to tailor criteria and performance levels. They then applied these rubrics to provide feedback on a sample block-based programming activity, followed by using a chatbot to deliver rubric-based feedback for the same work. Data were collected through Pre- and post-workshop surveys, open discussions and exit tickets. We used Thematic analysis for qualitative data. Teachers reported that they rarely create rubrics from scratch because the process is time-consuming and defining clear performance-level distinctions is challenging. After hands-on use, teachers described AI-generated rubrics as strong starting drafts that improved structure and clarified vague criteria, but emphasized the need for teacher oversight due to generic or grade-misaligned language, occasional misalignment with instructional priorities, and substantial editing requirements. Survey results indicated high perceived clarity and ethical acceptability, moderate assignment alignment, and usability as the primary weakness, particularly the ability to add, remove, or revise criteria. Open-ended responses highlighted a “strictness-versus-detail” tradeoff: AI feedback was often perceived as harsher but more detailed and scalable, leading teachers to express conditional willingness to adopt AI rubric tools when workflows support easy customization and preserve teacher control.

\end{abstract}

\section{Introduction}
The rapid growth in computer science enrollment, teachers face growing challenges in managing their classes effectively, making classroom management and individualized support increasingly demanding. To address these issues, digital tools and learning platforms have been introduced to assist teachers in supporting their tasks while also enhancing student learning outcomes and performance \cite{cockett2018use}. 
Rubrics are instruments that define criteria and performance standards, helping ensure consistency in grading while guiding both instructors and students in the assessment process \cite{english2022rubrics,jonsson2007use,vetrivel2025automated}. They offer teachers a systematic framework for evaluating student work and provide students with a clearer understanding of the standards expected of them \cite{allen2006rubrics}. By making assessment criteria explicit, rubrics allow students to see the rationale behind point deductions, enabling students to envision strategies for improving their work \cite{brookhart2015quality}. 
Providing effective and personalized feedback remains one of the most critical and time consuming responsibilities of educators. Due to heavy workloads of the teachers, the quality of feedback is often constrained \cite{liao2024feedbackpulse}. 
Recent advances in artificial intelligence (AI) have expanded these systems’ capabilities, enabling features such as automated rubric generation to streamline assessment practices \cite{omwando2025assessing}.
Unlike many existing autograders that require programming or specialized test-case design, rubric-based systems start from a familiar artifact that teachers already use for grading, making them more usable for novice K–12 teachers and aligned with existing practice \cite{ball2018lambda}. Once a rubric is defined, LLMs can leverage its structure to evaluate aspects of student work such as reasoning, creativity, and clarity, extending beyond what traditional code- or test-based autograders typically capture \cite{wang2021snapcheck}.

 
 Given the increasing ubiquity of AI in education, there is a growing consensus that both teachers and students must develop a foundational understanding of AI tools and their affordances. Equally, the design of effective AI systems requires careful alignment with users’ needs \cite{riahi2025comparative}, preferences \cite{riahi2021aesthetics}, perceptions and experience. This study examines the rubric generation feature of AI-enabled educational platforms, with particular attention to how it can enhance assessment and feedback practices in teaching and learning contexts.

The present study aims to: Examine the potential of artificial intelligence (AI) tools for rubric generation. Investigate teachers’ experiences, perceptions, and practices related to the use of these AI tools \cite{brookhart2013create}.
We conducted professional development sessions with middle- (n = 19) and high-school teachers (n = 6). During these sessions, teachers created rubrics with AI tool, and used the AI-generated rubrics to assess assignments. We gather teachers' feedback on the AI-enabled grading and rubric interface.
This study aim to answer the following research questions:

\begin{itemize}
   
    \item \textbf{RQ1.} What are teachers' perspectives about using a Large Language Model (LLM) for generating rubrics and making assessments for block-based programming in their classrooms?
     \item \textbf{RQ2.}How does hands-on exposure to AI rubric/assessment tools influence teachers’ intentions to adopt them in future classroom practice?
\end{itemize}

\section{Background}


Rubrics provide a coherent framework that spells out the criteria for student work and the quality levels for each, outlining the task’s key components, standards for success, and performance-level descriptors \cite{brookhart2013create,stevens2023introduction}. As detailed, consistent, requirement-aligned rubrics are essential for fair, transparent evaluation, recent work explores how AI can strengthen rubric development, improving precision, efficiency, and curricular alignment \cite{coronado2024enhancing}. AI can process rich, context-specific inputs and support iterative refinement, streamlining assessment design and enabling personalization \cite{fernandez2025navigating,tabarsi2025merryquery}.

Within CS education, research suggests that educational generative AI tools can be useful and well-received for assessment-adjacent support when paired with guardrails and instructor control \cite{liu2024teaching}. Related work on rubric-based AI assistance for grading similarly indicates that AI can generate reasonably accurate first-pass, rubric-aligned judgments that instructors validate, improving grading efficiency while keeping teachers in control; one study reports up to 85\% agreement with teacher criterion-level decisions in a CS course exam setting \cite{de2025autograde}.

Work in block-based learning environments has also examined AI-enabled assessment support, but often from a learner-facing perspective. For example, LevelUp is an automatic assessment tool embedded in a block-based machine learning environment that provides continuous, rubric-based formative feedback while learners are coding, and reports positive student perceptions of usefulness after the study \cite{reddy2022levelup,chidharom2024similar}. These systems emphasize in-the-moment scaffolding to improve student project quality during creation, rather than supporting teachers in authoring and adapting assessment criteria.

\begin{figure}[ht]
  \centering
  \includegraphics[width=0.6\textwidth]{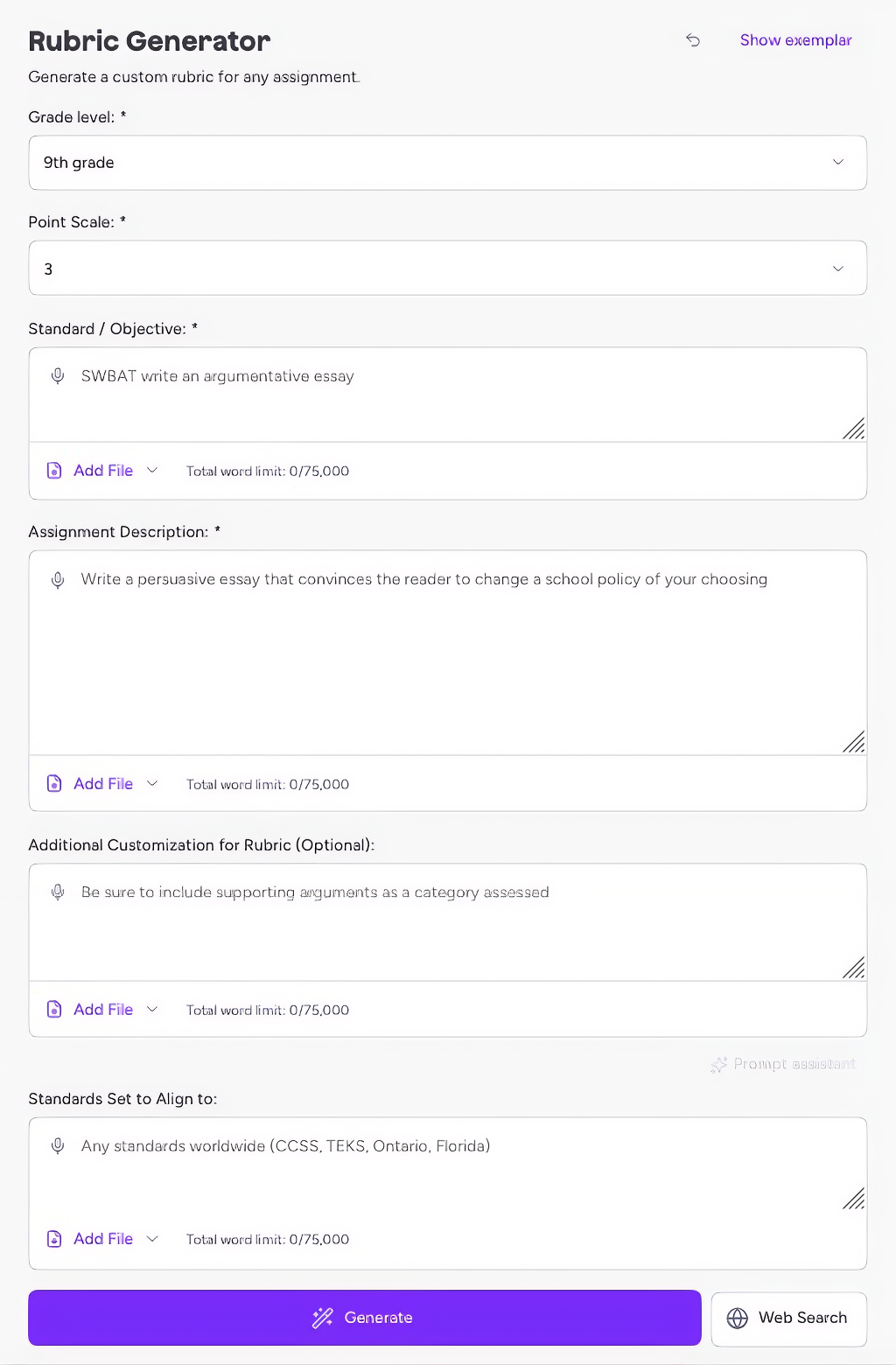}
  \caption{Magicschool.AI Rubric generation Tool}
  \label{fig:magic}
\end{figure}

Another instructor-facing tool that automates rubric creation from teacher inputs is Colleague AI’s Rubric Generation. It supports generating standards-aligned, editable rubrics from information such as grade level, subject, standards, goals, and uploaded materials, and it can also generate accompanying answer keys and feedback—reducing rubric authoring time and improving consistency while keeping teachers responsible for final grading decisions \cite{tian2025rubric}. This workflow is similar to MagicSchool.ai, which likewise prompts teachers to specify key context (e.g., grade level and scoring scale) and to provide assignment and standards information.
As shown in Figure~\ref{fig:magic}, MagicSchool.ai allows teachers to enter the grade level, scale/point structure, standards or objectives, and an assignment description, and it also provides an “additional customization” field for prompting specific rubric preferences (e.g., focus areas or tone). These shared design elements reflect a common pattern across instructor-facing rubric tools: the system generates an initial draft, while teachers retain control to adapt criteria, wording, and emphasis to match their instructional intent and classroom context.


More broadly, studies suggest AI-assisted rubrics may promote fairness and consistency by enforcing common criteria, tying judgments to explicit outcomes, and enabling timely individualized feedback, while also raising requests for greater transparency and explainability that underscore the importance of human oversight and iterative revision \cite{kanchana2025evolving}. Complementarily, comparisons of AI-generation feedback by chatbots with instructors and peers show that chatbots may provide detailed feedback but can be inconsistent or misaligned with human judgment, suggesting value in hybrid approaches that combine AI efficiency with human contextualization \cite{usher2025generative,tithi2025promise}.

Our work focuses on teacher-facing support for rubric generation and grading preparation in K–12 block-based programming contexts. We examine how teachers evaluate AI-generated rubrics and AI-assisted grading in terms of clarity, alignment, usability/editability, concerns \cite{atashpanjeh2022intermediate}, equity-related considerations, and how hands-on exposure during professional development shapes teachers’ intentions to adopt these tools in their classrooms.

\section{Method}

In Summer 2025, we conducted PD workshops with middle- and high-school teachers to examine their perceptions regarding AI-generated rubrics (Appendix \ref{app:first}). Teachers first reflected on their prior use of rubrics and any experience with AI via pre-survey (Table \ref{tab:combined_teacher_data}). After introducing an AI-rubric generator (magicschool.ai), teachers created a manual rubric for a sample coding task, then were introduced to North Carolina Department of Public Instruction (NCDPI) \cite{ncdpi2021cs} standards to craft a prompt and generate an AI rubric for comparison. They executed the sample code on snap.berkeley.edu and assessed it twice using the AI-generated rubric with and without AI chatbot in magicschool which name is Reina \cite{raina2025}. Teachers then completed follow-up questions via post-survey (Table \ref{tab:post}) about their experiences, concerns, and interest in future use. The two workshop versions for middle- and high-school differed only in task selection and scaffolding level.

\section{Qualitative Analysis}
We employed thematic analysis \cite{fereday2006demonstrating,maguire2017doing,alhojailan2012thematic} to examine qualitative data. Two coders independently analyzed the data, identified sub-themes, and the main researcher reviewed and refined the final themes (Appendix \ref{app:seccond}).
Across our analysis, eight themes emerged illustrating how teachers approach rubric creation and their experience with the AI-generated rubric tool interface. The themes were: \textbf{First}, teachers used rubrics for clarity and consistency, embedding them into platforms like Google Classroom and Canvas, though students sometimes struggled with text-heavy formats. \textbf{Second}, rubric design was time-consuming and challenging, particularly distinguishing middle performance levels and writing precise descriptors for new or creative tasks. \textbf{Third}, prior AI use centered on lesson planning, with few teachers applying AI to rubric creation. \textbf{Fourth}, AI-generated rubrics served as strong starting points, clarifying vague criteria, structuring creative tasks, and reducing workload. \textbf{Fifth}, perceptions of AI were mixed: teachers valued its speed and clarity but noted generic language, misalignment with objectives, and the need for substantial editing. \textbf{Sixth}, teachers raised concerns about fairness, equity, accuracy, privacy, and workflow barriers. \textbf{Seventh}, they recommended improvements such as LMS integration, better editing capabilities, grade-appropriate vocabulary, and scaffolded prompting. \textbf{Finally}, despite concerns, most teachers indicated they would use AI-rubric tools because they saved time while preserving teacher control over assessment design.

we analyzed every groups data to extract commonalities to shape sub-themes and themes.
\subsection{Current Rubric Practices}
Teachers across all groups described established routines for creating and using rubrics. Group 1 and Group 2, for example, use rubrics for nearly all assignments—including projects, math coursework, scientific experiments, and hands-on activities to maintain consistency and clarity. Group 3 emphasized the role of rubrics in helping students visualize expectations during multi-step or creative tasks. Group 4 and Group 5 teachers relied on rubrics embedded within Google Classroom, Canvas, or Facilita, although Group 1 noted that technical issues sometimes required printing rubrics instead of distributing them digitally. Several teachers (primarily Group 2 and Group 3) also used student self-assessments to complement rubric-based grading.

\subsection{Challenges in Creating Rubrics}
All groups described rubric creation as time-consuming and cognitively demanding, but the nature of the challenges varied. Groups 1, 3, and 5 highlighted the difficulty of distinguishing middle performance levels (e.g., scoring between “2,” “3,” and “4”), while Groups 2 and 4 struggled with writing precise descriptors that matched learning objectives. Teachers in Group 3 specifically noted that students often had trouble interpreting text-only rubrics. Groups 1 and 5 described practical issues such as students not accessing digital rubrics or LMS limitations. Groups 2 and 4 also expressed uncertainty about where AI should fit within their assessment workflow, indicating a need for clearer integration guidance.

\subsection{Use of AI for Rubric Creation}
AI use varied widely across groups. Group 2 and Group 5 used AI tools like ChatGPT to help structure creative or abstract assignments, particularly when criteria felt vague. Teachers in Group 4 reported relying on AI to generate initial drafts because they “dislike creating rubrics from scratch.” Group 1 participants used AI to interpret assignment documents and extract initial categories. In contrast, Group 3 teachers were familiar with AI for lesson planning but had limited experience applying it to rubric design. Across all groups, AI-generated rubrics were described as helpful starting points that required teacher refinement.

\subsection{Perceptions of AI-Generated Rubrics}
Positive perceptions were consistent but varied in emphasis. Group 1 participants were “happy and amazed” by the polished quality of AI-generated rubrics. Group 4 valued the speed and structure AI provided. Group 5 highlighted clearer descriptions and better alignment across performance levels, especially for new projects. Groups 2 and 3 emphasized AI’s strengths in clarifying vague criteria and helping personalize rubrics for student needs.
Negative perceptions clustered around similar limitations. Group 2 and Group 5 noted that AI language was often generic or not grade-appropriate. Group 4 pointed out that AI outputs still required significant editing, while Group 1 raised concerns about the difficulty level of certain AI-generated descriptors. Group 5 teachers worried that overreliance on AI could weaken their own assessment-design skills.
\subsection {Concerns and Risks}
Concerns were most strongly voiced by Groups 2 and 5. These included fairness, equity, and accuracy concerns, particularly whether AI could fully grasp learning objectives. Group 1 expressed discomfort with giving AI-generated rubrics directly to students when assignment outcomes were not yet clearly defined. Group 3 and Group 5 raised concerns about transparency, particularly whether students would understand how AI-generated criteria were produced. Privacy concerns were primarily mentioned by Group 2. Workflow constraints (e.g., AI blocked at school or limited LMS integration) were emphasized by Group 5.
\subsection{Conditions for Effective AI Use}

Group differences emerged clearly. Group 1 and Group 2 emphasized that AI is most effective for experienced teachers who can refine its outputs, while novices may overly depend on copy-and-paste results. Group 3 highlighted AI’s role in offering starting points but stressed the need for teacher oversight. Group 4 expressed confusion about how to use specific tool fields (e.g., “objective” vs. “description”), indicating the need for clearer scaffolding. Group 5 noted that teachers with limited subject knowledge may benefit the most from AI support.
\subsection{Recommendations for AI Tool Improvement}
Teachers requested improvements that spanned practical features and pedagogical alignment. Group 5 recommended better LMS integration, improved export options, and grade-level vocabulary support. Group 4 asked for clearer interfaces and better editability, particularly being able to adjust point distributions without regenerating the rubric. Group 1 suggested enabling AI to interpret assignment files (e.g., XML or uploads) more intelligently. Group 2 emphasized the need for scaffolded prompts and clearer separation of learning objectives and rubric descriptions.

\subsection{Future Intentions to Use AI}
Across all groups, teachers expressed interest in using AI rubric tools in their classrooms. Group 1 praised the tool’s flexibility and ability to handle file uploads and custom language. Group 2 and Group 4 appreciated the time savings and clarity provided by AI-generated rubrics. Group 3 valued AI’s ability to restore creativity to their workflow. Group 5 highlighted features such as multi-upload, editable outputs, and alignment with standards as key motivators for continued use. Across all groups, teachers stressed the importance of maintaining teacher control and reviewing AI outputs before using them with students.

\begin{table}[ht]
\centering
\caption{Teacher Demographics and Experience (Pre-survey)}
\label{tab:combined_teacher_data}
\begin{tabular}{lrr}
\toprule
\textbf{Category} & \textbf{n} & \textbf{\%} \\
\midrule
\multicolumn{3}{l}{\textit{Computer Science Teaching Experience ($N=25$)}} \\
\midrule
No experience & 9 & 36\% \\
1--3 years    & 5 & 20\% \\
4--7 years    & 4 & 16\% \\
8--10 years   & 7 & 28\% \\
\midrule
\multicolumn{3}{l}{\textit{Grades Currently Taught ($N=25$*)}} \\
\midrule
6--8   & 19 & 76\% \\
9--10  & 2  & 8\%  \\
10--12 & 4  & 16\% \\
\midrule
\multicolumn{3}{l}{\textit{Block-based Programming (BBP) Experience ($N=22$)}} \\
\midrule
No (does not teach BBP) & 11 & 55\% \\
Yes, 1--3 years         & 5  & 25\% \\
Yes, 4--7 years         & 1  & 5\%  \\
Yes, 8--10 years        & 5  & 15\% \\
\bottomrule
\addlinespace
\multicolumn{3}{l}{\small *Note: Total $n$ for grades taught may exceed $N$ if teachers select multiple ranges.}
\end{tabular}
\end{table}

\begin{figure}[ht]
  \centering
  \includegraphics[width=0.8\textwidth]{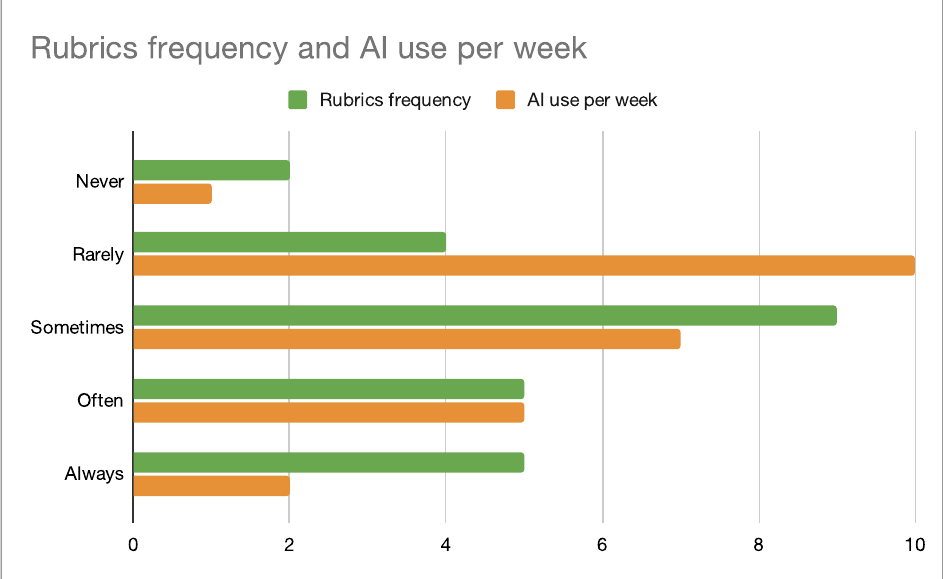}
  \caption{Pre-survey questions comparing level of using Rubrics and AI }
  \label{fig:pre2}
\end{figure}

\section{Survey Analysis}
In the pre-survey, we collected demographic information and prior experience teaching computer science (Table~\ref{tab:combined_teacher_data}). We also asked about their experience teaching block-based programming and the frequency with which they use AI and rubrics in their classes (Figure~\ref{fig:pre2}). The post-survey focused on teachers’ experiences using an AI-based rubric generation tool (Table~\ref{tab:post}). These items assessed perceived precision in representing performance levels; whether the language of the generated rubric was appropriate for students’ grade level; and the extent to which the rubric aligned with the curriculum, the assignment’s goals, and intended learning outcomes. We also asked whether the generated rubric captured the assessment focus areas teachers intended to grade. To evaluate usability, we included questions about how easily teachers could revise the rubric through iterative prompting, as well as the ease of adding or removing criteria. In addition, we measured equity-related perceptions, including whether the rubric was appropriate for students with different skill levels, whether teachers perceived potential bias or unfairness, and their confidence in and trust when using AI-generated rubrics with students. We further assessed perceived usefulness, such as whether the rubric supported students’ understanding of how their work would be evaluated and whether it promoted transparency in grading.

\subsection{Post-Survey Analysis}
Post-survey items measured teachers’ perceptions of the AI-generated rubric using a 5-point Likert scale (Strongly Agree to Strongly Disagree). Responses were converted to numeric scores (5 = Strongly Agree, 1 = Strongly Disagree). For reverse-coded items, we applied inverse numeric coding so that higher values consistently indicate more positive perceptions.

\begin{longtable}[c]{p{.15\linewidth} p{.3\linewidth} p{.55\linewidth} }
\caption{Post-survey items measured on a 5-point Likert scale.}
\label{tab:post}\\
\hline
\rowcolor[HTML]{DAE8FC} 
\textbf{Category} & \textbf{Sub-construct} & \textbf{Survey item} \\ \hline
\endhead
\hline
\endfoot
\endlastfoot
 & Criteria clarity & Q1: The criteria in the AI-generated rubric were clearly written. \\
 & Performance-level differentiation & Q2: The difference between levels of performance (e.g., Excellent, Good, Fair) were clear and easy to understand. \\
\multirow{-3}{=}{Clarity and readability} & Grade-level language appropriateness & Q3: The rubric language (vocabulary and tone) was suitable for the age or grade of the students. \\ \hline
 & Assignment goal alignment & Q4: The rubric aligned well with the goals of the assignment. \\
 & Curriculum / learning outcomes alignment & Q5: The AI-generated rubric captured broader teaching objectives, such as curriculum standards or learning outcomes. \\
 & Assessment focus fit & Q6: The rubric included the right focus areas (e.g., creativity, accuracy, reasoning) that the teacher wanted to grade. \\
\multirow{-4}{=}{Usefulness and alignment} & Student understanding of evaluation & Q7: The rubric could help students understand how their work will be evaluated. \\ \hline
 & Modifiability for classroom needs & Q8: I was easily able to modify the AI-generated rubric to suit my classroom needs. \\
 & Editing criteria control & Q9: The tool allowed me to add, remove, or revise rubric criteria without difficulty. \\
\multirow{-3}{=}{Usability and flexibility} & Flexible starting point & Q10: The rubric felt like a flexible starting point rather than a fixed product. \\ \hline
 & Perceived bias/ unfairness (absence of) & Q11: I did not see any indicators of potential bias or unfairness in the rubric. \\
 & Equitable use across learning needs & Q12: The rubric could be used equitably for students with different learning needs. \\
 & Grading transparency & Q13: The rubric promotes transparency in grading. \\
\multirow{-4}{=}{Ethical Factors} & Teacher confidence & Q14: I would feel confident using AI generated rubric with my students. \\ \hline
\end{longtable}

The items were organized into four constructs, and responses were analyzed      at the item level by computing the mean score for each question as follow:
\begin{itemize}
    \item \textbf{Clarity.} Overall, teachers reported positive perceptions of rubric clarity. Participants agreed that the criteria in the AI-generated rubric were clearly written (M = 3.9375, SD = 0.928) and that the difference between levels of performance (e.g., Excellent, Good, Fair) was clear and easy to understand (M = 4.125, SD = 0.957). Ratings were comparatively lower for whether the rubric language (vocabulary and tone) was suitable for the age or grade of the students (M = 3.5, SD = 1.211), suggesting that while criteria and performance-level distinctions were generally understandable, teachers were more mixed about grade-level appropriateness of wording.
\end{itemize}

\begin{itemize}
    \item \textbf{Usefulness and alignment.} Teachers generally perceived the AI-generated rubric as useful and reasonably aligned with instructional intent. Participants agreed that the rubric aligned well with the goals of the assignment (M = 4.0, SD = 1.264) and that the AI-generated rubric captured broader teaching objectives, such as curriculum standards or learning outcomes, to a moderate extent (M = 3.5, SD = 1.032). Ratings were also positive for whether the rubric included the right focus areas (e.g., creativity, accuracy, reasoning) that the teacher wanted to grade (M = 3.81, SD = 0.91). Participants further agreed that the rubric could help students understand how their work will be evaluated (M = 4.12, SD= 0.95), indicating strong perceived value for communicating expectations.
\end{itemize}

\begin{itemize}
    \item \textbf{Usability and flexibility.} Teachers’ responses indicated moderate usability and flexibility, with specific friction around editing criteria. Participants somewhat agreed that they were easily able to modify the AI-generated rubric to suit their classroom needs (M = 3.5, SD = 1.26) and that the rubric felt like a flexible starting point rather than a fixed product (M = 3.5625, SD = 1.31). In contrast, the lowest mean in this category was for the tool’s support for adding, removing, or revising rubric criteria without difficulty (M = 2.75, SD = 1.23), indicating a key usability limitation.
\end{itemize}

\begin{itemize}
    \item \textbf{Ethical factors.} Teachers generally agreed that they did not see indicators of potential bias or unfairness in the rubric (M = 3.93, SD = 1.062) and that the rubric promotes transparency in grading (M = 4.062, SD = 0.99). Confidence in using AI-generated rubrics with students was also positive (M = 3.87, SD = 1.024). The lowest rating in this category concerned whether the rubric could be used equitably for students with different learning needs (M = 3.125, SD = 1.147), suggesting greater uncertainty about how well the rubric generalizes across diverse learners.
\end{itemize}

\begin{figure}[ht]
  \centering
  \includegraphics[width=\textwidth]{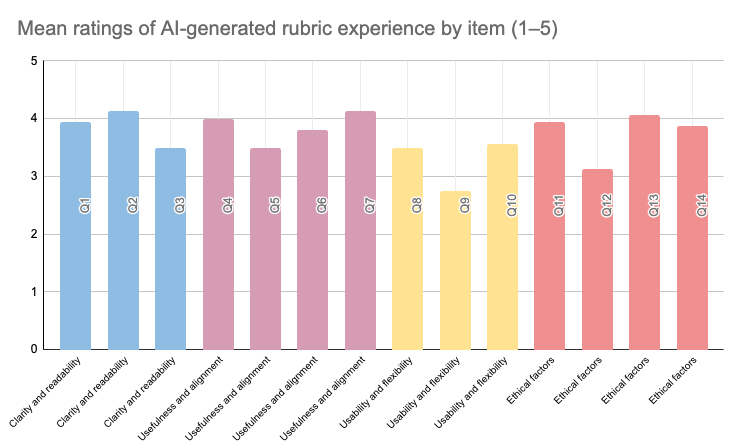}
  \caption{Post-Survey and Mean ratings of AI-generated rubric experience per item (1–5)}
  \label{fig:post}
\end{figure}
\subsection{Open-ended questions}
In the final segment of the professional development sessions, we included open-ended questions and exit ticket to better understand \textit{why teachers’ intentions to use AI platforms for rubric generation changed after hands-on exposure to these tools.}  We then asked teachers to rate the AI system’s rubric generation and feedback quality, describe their reactions to the AI-generated outputs, and reflect on whether and how they would incorporate these tools in their future classroom practice. Finally, we prompted participants to explain any shifts in their intention to use AI tools following the session and what factors contributed to that change.

Across responses about \textbf{how AI-generated feedback differed from teachers’ own feedback}, participants most often highlighted a trade-off between strictness and detail. Several teachers perceived the AI as harsher than they would be when assigning scores, while also noting that it produced more comprehensive, explanatory feedback than their typical comments (which were sometimes more numeric or brief). Teachers valued this added detail as potentially helpful for supporting large numbers of students, but they also emphasized the need for teacher oversight rather than full dependence on the AI. Others pointed to rubric interpretation and weighting issues (e.g., an “optional” feature being treated as a primary scoring criterion) and described how regenerating the rubric could improve alignment; they also requested more control over the type of summary and the granularity of the feedback.

When asked about the \textbf{depth of the AI-generated rubric and feedback}, participants generally rated it as substantial, with some describing it as near the top end of the scale due to its specificity and actionable suggestions. Teachers noted that prompt clarity and detail shaped the quality of the output, and that more descriptive prompts yielded deeper feedback. At the same time, some expressed concern that overly strict grading could discourage students, suggesting the system should allow teachers to calibrate strictness. Others rated the depth as moderately high (e.g., around 7–8/10), appreciating that the AI referenced concrete code structures and systematic implementation, while recommending improvements such as adding general programming best-practice feedback (e.g., naming conventions) beyond rubric-bound criteria.

Regarding overall experience and future use, teachers described the platform as easy to use and expected it to become more useful with trial-and-error and repeated practice, as they learned how to prompt effectively and refine outputs. Many indicated an intention to use AI-generated rubrics and feedback in future teaching, particularly because it could function as a clear communication artifact that helps students understand evaluation expectations. However, adoption was often framed as conditional: some wanted to test the tool more thoroughly on lower-quality or incorrect work and in realistic classroom scenarios before relying on it, while others raised practical constraints such as district restrictions that might require use outside of school contexts. Concerns about consistency of grading at scale also surfaced, reinforcing the view that AI support is promising but should remain configurable and teacher-directed.

\section{Discussions}

This study aims to not only evaluate perceived rubric quality, but also to understand how teachers reason about alignment with instructional intent, practical fit within their grading workflows, and equity-related implications when AI is involved in assessment design. Guided by this aim, we addressed two research questions focused on (1) teachers’ perceptions of AI-generated rubrics and (2) how exposure to such tools influences teachers’ intentions to use them in future teaching practice.

\textbf{RQ1: How do teachers perceive AI-based tools for rubric generation in block-based programming contexts?}

Teachers generally perceived AI-generated rubrics as a helpful starting point for assessment design, particularly because the tool could quickly produce a structured rubric with criteria and performance levels. Teachers described value in reducing the effort of drafting rubrics from scratch and in generating wording that can help formalize expectations. They also highlighted how AI-generated rubrics can support more consistent grading and clearer communication of expectations to students, which can improve transparency in how work is evaluated.

At the same time, teachers emphasized that AI-generated rubrics require teacher oversight and refinement. They reported that the tool sometimes produced rubrics that did not fully reflect their instructional priorities, such as emphasizing criteria differently than intended or missing important aspects of what they wanted to grade. Teachers also noted that the language and framing of criteria may need adjustment to better match students’ grade level and classroom norms. Overall, teachers positioned AI as an assistive resource that can accelerate rubric creation, but not as a substitute for teacher judgment in defining learning goals, selecting criteria, and ensuring the rubric fits the specific classroom context.

\textbf{RQ2: How does hands-on exposure to AI rubric/assessment tools influence teachers’ intentions to adopt them in future classroom practice?}

Hands-on exposure increased teachers’ ability to evaluate AI rubric tools realistically and helped them move from abstract opinions to concrete adoption considerations. After using the tool, many teachers reported greater openness to incorporating AI for rubric generation, especially when they could see how an AI-generated rubric could be iteratively refined and used to communicate expectations to students. The activity of applying the rubric to a sample assignment and comparing human grading with the tool’s grading outputs also helped teachers articulate what they would and would not trust the tool to do in practice.

However, teachers’ intentions to adopt these tools were often conditional. Teachers indicated they would be more likely to use AI rubric generation when the platform supports efficient customization, makes it easy to revise criteria and performance descriptions, and allows teachers to retain control over what is being assessed. They also discussed practical and ethical considerations that shape adoption, including whether the rubric can be applied fairly across students with different skill levels or learning needs, whether the tool could introduce unintended bias, and whether school or district policies allow classroom use \cite{zafarmand2025effect}. In short, exposure increased intention primarily when teachers could envision the tool fitting into a teacher-directed workflow where AI produces drafts and teachers finalize assessment decisions.

\textbf{Overall implication}
Across both research questions, the findings suggest that AI rubric tools are most likely to be adopted when they are designed to support teacher agency: enabling iterative refinement, alignment with goals and curriculum, and flexible adaptation for diverse learners, rather than presenting AI outputs as fixed or fully automated assessment products.

\section{Conclusion}
Our work contribute to investigating K–12 middle and high school teachers’ experiences with AI-supported rubric generation for block-based programming through professional development (PD) workshop sessions held between June and July 2025. We triangulated evidence from pre-survey context, post-survey evaluations, and open-ended reflections. The pre-survey captured teachers’ demographics and prior experience teaching computer science and block-based programming, as well as how frequently they used rubrics and AI in their classrooms. These responses showed that teachers entered the workshop with varied familiarity with AI tools and different grading routines, underscoring the value of structured, classroom-relevant exposure that enables teachers to evaluate AI tools in practice. After hands-on use of MagicSchool.ai to generate a rubric and Reina to generate feedback, teachers reported greater willingness to adopt AI rubric tools than at baseline, reflecting increased confidence in usability and effectiveness. Teachers particularly appreciated the clarity and structure of the generated rubrics mentioning these rubrics could support grading and help students understand evaluation expectations. They also viewed the rubrics as flexible and transparent, with few perceived bias concerns, increasing comfort with AI-assisted grading.

Teachers also highlighted key barriers to classroom use, especially the need for easier rubric editing and stronger support for equitable use across diverse learners. Although they saw alignment potential, they emphasized that teacher oversight remains necessary to ensure criteria, weighting, and language fit the assignment and grade level. Open-ended responses clarified why teachers’ intentions shifted after the PD experience. Teachers treated AI-generated rubrics as time-saving drafts and stressed the need for teacher control to ensure alignment with assignment goals, grade-level language, and intended focus areas; when comparing their grading to the platform’s automated outputs, they valued scalable, detailed feedback but noted moments of strictness or misalignment. Together, these reflections suggest AI rubric tools should prioritize teacher-directed customization rather than fixed automation.

The PD workshop itself also influenced adoption intentions by offering a guided setting to try the tools, reflect, and form more grounded judgments about usefulness and usability. Despite ongoing concerns—especially about editing flexibility and equitable application—teachers remained interested in using these tools, indicating that well-designed PD can support more informed and confident adoption.

\section{Future work}
Our findings point to several directions for future work on AI rubric generation in K–12 settings. Next-generation tools should better support teacher agency through stronger editing and versioning workflows, clearer mechanisms for aligning rubrics with standards and learning outcomes, and built-in scaffolds for promoting transparency and equitable use across diverse learners. Future research should also examine how these design improvements affect teachers’ sustained adoption and students’ assessment experiences in real classroom deployments.

\begin{credits}
\subsubsection{\ackname} This material is based upon work supported by the National Science Foundation under Award No. 2405854 and 2405855. Any opinions, findings and conclusions, or recommendations expressed in this material are those of the authors and do not necessarily reflect the views of the sponsor.

\end{credits}

%
%
%

 \bibliographystyle{splncs04}
 \bibliography{sample-base}

@article{cockett2018use,
  title={The use of assessment rubrics to enhance feedback in higher education: An integrative literature review},
  author={Cockett, Andrea and Jackson, Carole},
  journal={Nurse education today},
  volume={69},
  pages={8--13},
  year={2018},
  publisher={Elsevier}
}

@article{english2022rubrics,
  title={Rubrics and formative assessment in K-12 education: A scoping review of literature},
  author={English, Narelle and Robertson, Pam and Gillis, Shelley and Graham, Lorraine},
  journal={International Journal of Educational Research},
  volume={113},
  pages={101964},
  year={2022},
  publisher={Elsevier}
}

@article{jonsson2007use,
  title={The use of scoring rubrics: Reliability, validity and educational consequences},
  author={Jonsson, Anders and Svingby, Gunilla},
  journal={Educational research review},
  volume={2},
  number={2},
  pages={130--144},
  year={2007},
  publisher={Elsevier}
}

@incollection{vetrivel2025automated,
  title={Automated Grading Systems: Enhancing Efficiency and Consistency in Student Assessments},
  author={Vetrivel, SC and Arun, VP and Ambikapathi, Ramya and Saravanan, TP},
  booktitle={Adopting Artificial Intelligence Tools in Higher Education},
  pages={41--61},
  year={2025},
  publisher={CRC Press},
  address = {Boca Raton}
}

@article{allen2006rubrics,
  title={Rubrics: Tools for making learning goals and evaluation criteria explicit for both teachers and learners},
  author={Allen, Deborah and Tanner, Kimberly},
  journal={CBE—Life Sciences Education},
  volume={5},
  number={3},
  pages={197--203},
  year={2006},
  publisher={American Society for Cell Biology}
}

@article{brookhart2015quality,
  title={The quality and effectiveness of descriptive rubrics},
  author={Brookhart, Susan M and Chen, Fei},
  journal={Educational Review},
  volume={67},
  number={3},
  pages={343--368},
  year={2015},
  publisher={Taylor \& Francis},
  address={Philadelphia, PA}
}

@inproceedings{liao2024feedbackpulse,
  title={FeedbackPulse: GPT-Enabled Feedback Assistant for Software Engineering Educators},
  author={Liao, Yiwen and Jiang, Yuchao and Chen, Zhangpeng and Suleiman, Basem},
  booktitle={2024 36th International Conference on Software Engineering Education and Training (CSEE\&T)},
  pages={1--2},
  year={2024},
  publisher={IEEE},
  address={Piscataway, New Jersery}
}

@inproceedings{omwando2025assessing,
  title={Assessing the Impact of the Use of Generative AI in Developing and Using Assessment Grading Rubrics for Engineering Courses},
  author={OMWANDO, THOMAS AMING'A and Alhalawani, Adel and Khandha, Ashutosh and Kotla, Bhavana},
  booktitle={2025 ASEE Annual Conference \& Exposition},
  year={2025},
  numpages={46},
  publisher = {ASEE},
  address={Quebec, Canda}
}

@book{brookhart2013create,
  title={How to create and use rubrics for formative assessment and grading},
  author={Brookhart, Susan M},
  year={2013},
  publisher={Ascd},
  address={Alexandria, VA}
}

@book{stevens2023introduction,
  title={Introduction to rubrics: An assessment tool to save grading time, convey effective feedback, and promote student learning},
  author={Stevens, Dannelle D},
  year={2023},
  publisher={Routledge},
  address={New York, NY}
}

@inproceedings{coronado2024enhancing,
  title={Enhancing Educational Innovation: Evaluating the Accuracy of AI-Generated Assessments for Engineering Course Reports},
  author={Coronado-Apodaca, Karina G and Barrios-Pi{\~n}a, H{\'e}ctor A},
  booktitle={International Conference on Technology and Innovation in Learning, Teaching and Education},
  pages={90--104},
  year={2024},
  publisher={Springer},
  address={Cham, Switzerland}
}

@article{fernandez2025navigating,
  title={Navigating the future of pedagogy: The integration of AI tools in developing educational assessment rubrics},
  author={Fern{\'a}ndez-S{\'a}nchez, Andrea and Lorenzo-Casti{\~n}eiras, Juan Jos{\'e} and S{\'a}nchez-Bello, Ana},
  journal={European Journal of Education},
  volume={60},
  number={1},
  pages={e12826},
  year={2025},
  publisher={Wiley Online Library}
}

@article{tian2025rubric,
  title={Rubric Generation in Colleague AI: Transforming Assessment in Education},
  author={Tian, Zewei Victor and Esbenshade, Lief and Liu, Alex and Sarkar, Shawon and Zhang, Zachary and He, Kevin and Sun, Min},
  journal={Social Innovations Journal},
  volume={30},
  number={2},
  numpages={6},
  year={2025}
}

@article{kanchana2025evolving,
  title={Evolving Student Assessment: AI-Driven Rubrics for Personalized and Equitable English Language Learning},
  author={Kanchana, S and Saha, Prativa Rani},
  journal={Journal of Engineering Education Transformations},
  pages={584--590},
  year={2025},
  volume = {38}
}

@article{alhojailan2012thematic,
  title={Thematic Analysis: A Critical Review of Its Process and Evaluation},
  author={Alhojailan, Mohammad I.},
  journal={West East Journal of Social Sciences},
  month={12},
  volume = {1},
  number = {1},
  pages={9},
  year={2012},
  publisher={The West East Institute}
}

@article{maguire2017doing,
  title={Doing a Thematic Analysis: A Practical, Step-by-Step Guide for Learning and Teaching Scholars},
  author={Maguire, M. and Delahunt, B.},
  journal={All Ireland Journal of Higher Education},
  volume={9},
  number={3},
  pages={},
  year={2017}
}

@article{fereday2006demonstrating,
  title={Demonstrating rigor using thematic analysis: A hybrid approach of inductive and deductive coding and theme development},
  author={Fereday, Jennifer and Muir-Cochrane, Eimear},
  journal={International journal of qualitative methods},
  volume={5},
  number={1},
  pages={80--92},
  year={2006},
  publisher={SAGE Publications Sage CA: Los Angeles, CA}
}

@misc{ncdpi2021cs,
  author       = {North Carolina Department of Public Instruction},
  title        = {K--12 Computer Science Standards},
  year         = {2021},
  publisher    = {NCDPI},
  address      = {Raleigh, NC},
  note         = {Retrieved from https://www.dpi.nc.gov/}
}

@inproceedings{wang2021snapcheck,
  title={SnapCheck: Automated testing for snap! programs},
  author={Wang, Wengran and Zhang, Chenhao and Stahlbauer, Andreas and Fraser, Gordon and Price, Thomas},
  booktitle={Proceedings of the 26th ACM Conference on Innovation and Technology in Computer Science Education V. 1},
  pages={227--233},
  year={2021},
  publisher = {ACM},
  address={New York, NY}
}

@mastersthesis{ball2018lambda,
    Author= {Ball, Michael},
    Editor= {Garcia, Dan},
    Title= {Lambda: An Autograder for Snap!},
    School= {EECS Department, University of California, Berkeley},
    Year= {2018},
    Month= {Jan},
    Url= {http://www2.eecs.berkeley.edu/Pubs/TechRpts/2018/EECS-2018-2.html},
    Number= {UCB/EECS-2018-2}
}

@inproceedings{liu2024teaching,
  title={Teaching CS50 with AI: leveraging generative artificial intelligence in computer science education},
  author={Liu, Rongxin and Zenke, Carter and Liu, Charlie and Holmes, Andrew and Thornton, Patrick and Malan, David J},
  booktitle={Proceedings of the 55th ACM technical symposium on computer science education V. 1},
  pages={750--756},
  year={2024},
  publisher = {ACM},
  address={New York, NY}
}

@inproceedings{de2025autograde,
  title={" AutoGrade": an AI-Based Assessment Tool for Computer Science 1},
  author={de Kereki, In{\'e}s Friss and Garrido, Ismael},
  booktitle={2025 IEEE Engineering Education World Conference (EDUNINE)},
  pages={1--6},
  year={2025},
  publisher={IEEE},
address={Piscataway, New Jersery}
}

@article{usher2025generative,
  title={Generative AI vs. instructor vs. peer assessments: a comparison of grading and feedback in higher education},
  author={Usher, Maya},
  journal={Assessment \& Evaluation in Higher Education},
  pages={1--16},
  year={2025},
  publisher={Taylor \& Francis},
  volume={50},
  number={6}
}

@inproceedings{reddy2022levelup,
  title={LevelUp--automatic assessment of block-based machine learning projects for AI education},
  author={Reddy, Tejal and Williams, Randi and Breazeal, Cynthia},
  booktitle={2022 IEEE Symposium on Visual Languages and Human-Centric Computing (VL/HCC)},
  pages={1--8},
  year={2022},
  publisher={IEEE},
  address={Piscataway, New Jersery}
}

@article{chidharom2024similar,
  title={Similar Quality of Visual Working Memory Representations between Negative and Positive Attentional Templates.},
  author={Chidharom, M and Zafarmand, M and Carlisle, NB},
  journal={Journal of Cognition},
  volume={7},
  number={1},
  pages={55--55},
  year={2024}
}

@article{zafarmand2025effect,
  title={The Effect of Visual and Verbal Cueing on Semantic Bias Activation and Target Selection},
  author={Zafarmand, Mahsa and Zou, Xinying and Carlisle, Nancy},
  journal={Journal of Vision},
  volume={25},
  number={9},
  pages={2770--2770},
  year={2025},
  publisher={The Association for Research in Vision and Ophthalmology}
}

@inproceedings{atashpanjeh2022intermediate,
  title={Intermediate help with using digital devices and online accounts: understanding the needs, expectations, and vulnerabilities of young adults},
  author={Atashpanjeh, Hanieh and Behfar, Arezou and Haverkamp, Cassity and Verdoes, Maryellen McClain and Al-Ameen, Mahdi Nasrullah},
  booktitle={International Conference on Human-Computer Interaction},
  pages={3--15},
  year={2022},
  publisher={Springer},
  address={Cham, Switzerland}
}

@inproceedings{riahi2025comparative,
  title={Comparative Analysis of STEM and Non-STEM Teachers’ Needs for Integrating AI into Educational Environments},
  author={Riahi, Bahare and Catet{\'e}, Veronica},
  booktitle={International Conference on Human-Computer Interaction},
  pages={125--140},
  year={2025},
  publisher={Springer},
  address = {Cham, Switzerland}
}

@misc{tithi2025promise,
  title={The Promise and Limits of LLMs in Constructing Proofs and Hints for Logic Problems in Intelligent Tutoring Systems},
  author={Tithi, Sutapa Dey and Ramesh, Arun Kumar and DiMarco, Clara and Tian, Xiaoyi and Alam, Nazia and Fazeli, Kimia and Barnes, Tiffany},
  journal={arXiv preprint arXiv:2505.04736},
  year={2025}
}

@inproceedings{tabarsi2025merryquery,
  title={MerryQuery: A Trustworthy LLM-Powered Tool Providing Personalized Support for Educators and Students},
  author={Tabarsi, Benyamin and Basarkar, Aditya and Liu, Xukun and Xu, Dongkuan DK and Barnes, Tiffany},
  booktitle={Proceedings of the AAAI Conference on Artificial Intelligence},
  volume={39},
  pages={29700--29702},
  year={2025},
  publisher = {PKP},
  address={Burnaby, Canada}
}

@inproceedings{riahi2021aesthetics,
  title={Aesthetics for communicative visualization: A brief review},
  author={Riahi, Maryam and Watson, Benjamin},
  publisher={OSF},
  year = {2021},
  booktitle = {Proceedings of the 35th Visual Communications Conference (VisComm)},
  numpages ={6},
  address = {Washington, DC}
}

@misc{raina2025,
  author       = {{MagicSchool.ai}},
  title        = {Raina},
  year         = {2025},
  howpublished = {\url{https://app.magicschool.ai/raina}},
  note         = {AI chatbot}
}

\appendix

\section{Appendix 1}
\subsection{Magicschool AI Output of Rubric for Brick Wall Activity}
\label{app:first}

\small
\setlength{\tabcolsep}{3pt}
\renewcommand{\arraystretch}{1.1}

\begin{longtable}{|p{.19\linewidth}|p{.20\linewidth}|p{.20\linewidth}|p{.20\linewidth}|p{.21\linewidth}|}
\caption{Rubric for 9th-Grade Snap Programming: Brick Wall Abstraction Assignment.}
\label{tab:snap-rubric}\\
\hline
\rowcolor[HTML]{DAE8FC}
\textbf{Criteria} & \textbf{4 -- Excellent} & \textbf{3 -- Proficient} & \textbf{2 -- Developing} & \textbf{1 -- Beginning} \\
\hline
\endfirsthead

\hline
\rowcolor[HTML]{DAE8FC}
\textbf{Criteria} & \textbf{4 -- Excellent} & \textbf{3 -- Proficient} & \textbf{2 -- Developing} & \textbf{1 -- Beginning} \\
\hline
\endhead

\hline
\endfoot

\hline
\endlastfoot

Design of ``Draw Brick'' Block &
Custom block accurately draws a brick with correct color, flat line ends, and precise dimensions; handles half bricks appropriately. &
Brick block draws a brick mostly correctly but has minor inaccuracies in color, line ends, or size. &
Brick block is created but has noticeable errors in color, line ends, or size; half bricks missing or incorrect. &
Brick block is incomplete or missing; does not represent a brick or lacks correct features. \\
\hline

Problem Decomposition \& Abstraction &
Effectively breaks down the task into logical reusable blocks for bricks, rows (Row A and Row B), and full wall; demonstrates clear abstraction. &
Breaks down task into blocks with minor gaps in abstraction or reusability. &
Attempts decomposition but lacks clarity or reusability; some blocks do not function as intended. &
Does not decompose task into meaningful blocks; abstraction is missing or ineffective. \\
\hline

Handling Even/Odd Rows \& Mortar &
Correctly implements logic to handle both even and odd numbers of rows; mortar thickness and spacing between bricks and rows are precise and adjustable. &
Handles even/odd rows with minor logical errors; mortar thickness and spacing mostly consistent. &
Inconsistent or incorrect handling of row types; mortar thickness and spacing irregular or fixed. &
Does not account for row differences or mortar gaps; mortar and spacing missing or incorrect. \\
\hline

Parameterization \& Flexibility &
Wall length, brick size, and mortar thickness are all adjustable parameters; changes dynamically affect half bricks and row spacing. &
Most parameters adjustable; changes affect wall with minor issues in half-brick size or spacing. &
Limited parameter adjustments; changes inconsistently impact wall design. &
No parameterization; wall is fixed size and design without flexibility. \\
\hline

Code Reusability \& Logical Thinking &
Code blocks are reusable, logically organized, and clearly named; program structure shows strong understanding of abstraction concepts. &
Code is mostly reusable and organized with some naming or logic inconsistencies. &
Limited reuse of code blocks; program organization and naming lack clarity. &
Code is repetitive, disorganized, or lacks logical flow and abstraction principles. \\
\hline

Visual Accuracy of Brick Wall &
Brick wall visually matches design expectations with correct number of rows and bricks; bricks and mortar appear proportional and consistent. &
Brick wall mostly accurate with minor visual inconsistencies in brick count, size, or mortar. &
Brick wall is incomplete or visually inconsistent; noticeable errors in brick size or mortar gaps. &
Brick wall is incomplete, visually incorrect, or missing significant elements. \\
\hline

\end{longtable}

\subsection{Rubric Themes}
\label{app:seccond}



\begin{longtable}[c]{|p{.19\linewidth}|p{.23\linewidth}|p{.1\linewidth}|p{.55\linewidth}|}
\caption{Themes identified from teacher rubric creation interviews.}
\\
\hline
\rowcolor[HTML]{DAE8FC}
Theme & Subtheme & Groups & Summary of Findings \\ \hline
\endfirsthead

\hline
\rowcolor[HTML]{DAE8FC}
Theme & Subtheme & Groups & Summary of Findings \\ \hline
\endhead

\hline
\endfoot

\hline
\endlastfoot

\multirow{4}{=}{1. Current Rubric Practices} & Regular use of rubrics & G1, G2, G3, G4, G5 & Teachers consistently use rubrics for projects, math, labs, and hands-on activities; formats range from analytic scales to checklists. \\ \cline{2-4}
 & Integration into LMS & G1, G2, G5 & Rubrics created in Google Classroom, Canvas, Facilita; technical issues occasionally require paper copies. \\ \cline{2-4}
 & Rubrics guide learning \& grading & G1, G2, G3 & Rubrics communicate expectations and structure assessment; some teachers incorporate student self-assessment. \\ \cline{2-4}
 & Student use of rubrics & G2, G3 & Students appreciate rubrics but struggle with text-heavy or inaccessible digital formats. \\ \hline

\multirow{5}{=}{2. Challenges in Creating Rubrics} & Distinguishing performance levels & G1, G3, G5 & Teachers struggle to differentiate mid-range scores (e.g., 2 vs.\ 3 vs.\ 4). \\ \cline{2-4}
 & Writing clear descriptors & G2, G4, G5 & Difficult to craft precise, well-defined criteria, especially for new or creative tasks. \\ \cline{2-4}\cline{1-1}
 & Student interpretation challenges & G3, G5 & Students have difficulty visualizing expectations from text-only rubrics. \\ \cline{2-4}\cline{1-1}
 & Limited AI assessment knowledge & G2, G4, G5 & Teachers know AI for lesson planning but not for rubric creation or editing. \\ \cline{2-4}\cline{1-1}
 & Technical \& workflow issues & G1, G5 & Online sharing, saving, and exporting issues hinder rubric distribution. \\ \hline

\multirow{3}{=}{3. Use of AI for Rubric Creation} & AI as a starting point & G1, G2, G4, G5 & AI helps initiate rubric drafting, especially for vague or abstract criteria. \\ \cline{2-4}
 & Limited familiarity with AI rubrics & G3, G5 & AI mainly used for lesson prep; teachers unsure how to integrate AI rubrics. \\ \cline{2-4}\cline{1-1}
 & Reduces workload & G2, G4 & Teachers who dislike writing rubrics rely on AI for quick initial drafts. \\ \hline

\multirow{4}{=}{4. Perceptions of AI-Generated Rubrics \vspace{3mm}} & Positive perceptions & G1, G2, G3, G4, G5 & AI rubrics viewed as polished, fast to generate, and well-structured. \\ \cline{2-4}
 & Personalization \& clarity & G2, G3 & AI improves clarity of point distinctions and supports tailored criteria. \\ \cline{2-4}
 & Limitations \& misalignment & G2, G4, G5 & Outputs may be generic, misaligned with learning objectives, or require editing. \\ \cline{2-4}
 & Assignment mismatch & G1, G5 & Some rubrics did not match intended student tasks or expected outputs. \\ \hline

\multirow{4}{=}{5. Concerns and Risks} & Fairness, accuracy, equity & G2, G5 & Worries about mis-scoring, misunderstanding learning goals, or bias. \\ \cline{2-4}
 & Hesitation to give to students & G1, G5 & Concern that students won’t understand AI-generated rubrics or trust them. \\ \cline{2-4}
 & Privacy concerns & G2 & Fear of AI storing student information. \\ \cline{2-4}
 & Access \& workflow barriers & G2, G5 & Schools may block AI; LMS integration is incomplete. \\ \hline

\multirow{3}{=}{6. Conditions for Effective AI Use} & Best for experienced teachers & G1, G2, G4 & Experienced teachers refine AI output effectively; novices risk over-reliance. \\ \cline{2-4}
 & AI not as a final product & G2, G3, G4 & Teachers emphasize reviewing and aligning AI rubrics with objectives. \\ \cline{2-4}
 & Need for scaffolding & G2, G4, G5 & Teachers want clearer prompts, examples, and instructions within the tool. \\ \hline

\multirow{5}{=}{7. Recommendations for Improvement} & Better editability and scoring control & G4, G5 & Allow point adjustments without regenerating rubrics; better editing tools. \\ \cline{2-4}
 & Grade-level appropriate vocabulary & G2, G5 & Improve alignment with student reading levels and disciplinary standards. \\ \cline{2-4}
 & LMS integration \& exporting & G1, G5 & Direct syncing with Google Classroom/Canvas \& simpler export options. \\ \cline{2-4}
 & Prompt scaffolding for novices & G2, G4 & Clearer separation of objectives, descriptions, and criteria in the interface. \\ \cline{2-4}
 & Advanced features (e.g., file parsing) & G1 & Teachers want AI to interpret assignment documents automatically. \\ \hline

\multirow{3}{=}{8. Future Intentions to Use AI} & Likely to use AI rubrics & G1, G2, G4, G5 & Majority plan to adopt AI due to time-saving and helpful structure. \\ \cline{2-4}
 & Reasons for adoption & All groups & AI provides speed, organization, clarity, and personalizable criteria. \\ \cline{2-4}
 & Reservations & G2, G5 & Accuracy, school AI restrictions, and desire for teacher control. \\ \hline

\end{longtable}

\end{document}